\documentclass{aa}
\usepackage{txfonts}
\usepackage{color}
\usepackage{graphicx}
\usepackage{ulem}
\usepackage{cancel}
\usepackage{tikz}

\definecolor{dodgerblue}{rgb}{0.12,0.56,1.0}
\definecolor{indianred}{rgb}{0.8,0.36,0.36}
\definecolor{darkgreen}{rgb}{0.0,0.5,0.0}
\definecolor{gray}{rgb}{0.5,0.5,0.5}
\newcommand{\avg}[1]{\ensuremath{\langle #1 \rangle}}
\begin{document}

   \title{Small-scale dynamo in cool stars}
   \subtitle{I. Changes in stratification and near-surface convection for main-sequence spectral types}
    \author{Tanayveer S. Bhatia\inst{1}\fnmsep\thanks{\email{bhatia@mps.mpg.de}}\and
    Robert H. Cameron\inst{1}\and
    Sami K. Solanki\inst{1}\fnmsep\inst{2}\and
    Hardi Peter\inst{1}\and
    Damien Przybylski\inst{1}\and
    Veronika Witzke\inst{1}\and
    Alexander Shapiro\inst{1}}
    
    \institute{Max-Planck-Institut f\"ur Sonnensystemforschung, Justus-von-Liebig Weg 3, 37077 Göttingen
    \and
        School of Space Research, Kyung Hee University, Yongin, Gyeonggi-Do, 446-701, Korea
    }
   \date{Submitted: 22 Mar 2022, Accepted: 31 May 2022}
 
 \abstract
 {Some of the small-scale solar magnetic flux can be attributed to a small-scale dynamo (SSD) operating in the near-surface convection. The SSD fields have consequences for solar granular convection, basal flux, and chromospheric heating. A similar SSD mechanism is expected to be active in the near-surface convection of other cool main-sequence stars, but this has not been investigated thus far.}
 {We aim to investigate changes in stratification and convection due to inclusion of SSD fields for F3V, G2V, K0V, and M0V spectral types in the near-surface convection.}
 {We studied 3D magnetohydrodynamic (MHD) models of the four stellar boxes, covering the subsurface convection zone up to the lower photosphere in a small Cartesian box, based on the \textit{MURaM} radiative-MHD simulation code. We compared the SSD runs against reference hydrodynamic runs.}
 {The SSD is found to efficiently produce magnetic field with energies ranging between 5\% to 80\% of the plasma kinetic energy at different depths. This ratio tends to be larger for larger $T_{\mathrm{eff}}$. The relative change in density and gas pressure stratification for the deeper convective layers due to SSD magnetic fields is negligible, except for the F-star. For the F-star, there is a substantial reduction in convective velocities due to Lorentz force feedback from magnetic fields, which, in turn, reduces the turbulent pressure.}
 {The SSD in near-surface convection for cool main-sequence stars introduces small but significant changes in thermodynamic stratification (especially for the F-star) due to a reduction in the convective velocities.}

   \keywords{Stars: interiors --
                Stars: magnetic fields --
                Stars: late-type --
                Convection --
                Dynamo
               }

   \maketitle
%

\section{Introduction}\label{sec:intro}
The interpretation of data from stellar observations requires comparisons against stellar models. Traditionally, these models have been 1D global models \citep{carbon1969} that use formulations of mixing-length theory (MLT) \citep{bv_mixinglength}. Later models have accounted for line-blanketing effects and used an opacity distribution function (ODF) approach to calculate opacities \citep{strom1966}. Among them, the \textit{MARCS} code \citep{gustafsson1975,gustafsson2008}, the \textit{ATLAS} code \citep{kurucz1979,castelli2003}, and the \textit{PHOENIX} code \citep{phoenix1995,phoenix1999} have enabled the calculation of synthetic stellar spectra with a detailed accounting for the relevant physics. These models have enabled, for example, the accurate determination of abundances and stellar evolution tracks, along with constraining the chemical evolution of galaxies \citep{edvardsson1993}.

Nonetheless, convection is a 3D process and a phenomenological 1D approach is insufficient for characterizing the properties of granulation and plasma turbulence. Radiative aspects of granulation, in particular, are not fully captured by an MLT-like approach. The 1D models also require free parameters, such as the mixing length parameter, $\alpha_{MLT}$, and turbulent velocities for line broadening calculations. Hence, 3D stellar atmosphere models become important for a realistic interpretation of stellar characteristics from observations. The first 3D simulations of solar convection were pioneered by \citet{nordlund1982,stein1998,nordlund1990}. These simulations are realistic in the sense that they are directly comparable to solar observations: they reproduce granulation pattern and associated spectral line widths and asymmetries quite well \citep{asplund2000}.

Early stellar models \citep{nordlund1990granulation} showed the limitations of MLT-based models in accurately reproducing the near-surface temperature gradient, which affects radiative properties in the lower photospheres. Subsequently, various models have been constructed over a grid of effective temperature, surface gravity, and metallicity. Currently, the most comprehensive 3D grids \citep{stagger2013,cifist2009} cover a wide range of stellar type on the HR-diagram, but they are purely hydrodynamic. 

However, stellar convection is not a purely hydrodynamic process. Most cool stars are expected to have magnetic fields. Hence, a complete description of their photospheres should also take into account the effects of such fields. The best studied star in this context is the Sun. There is a rich variety of solar magnetic field-related phenomena ranging from sunspots and active regions to network fields, forming plages and faculae all the way down to small-scale mixed polarity turbulent magnetic field filling the rest of the solar surface. State-of-the-art solar simulations reproduce all of these features, from sunspots \citep{rempel_sunspot_2009} and plages \citep{VogSch2005,yadav_plage_2021} to quiet-sun magnetism \citep{rempel2014}.

The quiet-sun small-scale field, that is, the field associated with regions of the sun not showing any activity, is partly attributable to a small-scale dynamo (SSD) operating in the convection zone \citep{VogSch2007,rempel2014,GCS2010}. This field is believed to have a significant magnitude of around $\sim$ 130 G, based on Hanle depolarization \citep{trujillo2004}. Additional evidence that a fair fraction of the small-scale field is a result of an SSD comes from the fact that internetwork magnetic flux does not follow the solar cycle \citep{buehler2013,lites2014}. In realistic radiative-MHD simulations, the effect of quiet-sun magnetic fields (self consistently generated via an SSD mechanism) has previously been shown to be important, for example, to reproduce the correct solar intensity contrast \citep{danilovic2010} and account for inferred photospheric magnetic field strength based on Hanle-effect diagnostics \citep{shchukina2011}. In addition, there is a tendency to obtain a rough equipartition between kinetic and magnetic energy in SSD simulations \citep{hotta_ssd,haugen_ssd,schekochihin_ssd}, which implies a substantial reduction in plasma velocities since most of the energy in the magnetic fields is obtained from the plasma motions. The importance of magnetic fields generated from an SSD in other stellar types and its effect on the intensity characteristics, however, remains to be explored. Hence, it is imperative to investigate the effect of quiet-star small-scale magnetic fields on these quantities in a subsequent study.

This paper is a part of a project aimed at constructing a grid of magneto-convective stellar atmospheres, ranging across temperatures ($3500 < T < 7000$ K), gravity ($ 4.3 < \log_{10} g < 4.8$ in cgs units), and metallicities (in this paper, only solar metallicities are considered). We investigate four stellar cases: F3V, G2V, K0V, and M0V and we study the relative change in stratification, convection, and intensity from purely hydrodynamic setups.

In §\ref{sec:methods}, we outline the simulation code and the setup. Then, we present the results of the simulations in §\ref{sec:results}, followed by interpretation of the results in §\ref{sec:discuss}. Lastly, we summarize the results and present the corresponding discussion in §\ref{sec:conclude}.
\section{Methods}\label{sec:methods}
\subsection{Simulation code}\label{sec:methods:code}
The code we use throughout this work is \textit{MURaM} \citep{VogSch2005,rempel2014,rempel2017}, a 3D radiative-MHD code that solves the conservative MHD equations for compressible, partially ionized plasma. It uses a multi-group radiative transfer scheme with short characteristics \citep{nordlund1982}.
The equations for mass ($\rho$ - density), momentum ($\vec{v}$ - velocity, $p$ - pressure), and energy ($\epsilon_h$ - enthalpy density) conservation are solved, along with the induction equation ($\vec{B}$ - magnetic field)

\begin{align}
        \partial_{t}\rho=&-\nabla\cdot(\rho\vec{v}),\\
        \partial_{t}(\rho\vec{v})=&-\nabla\cdot(\rho\vec{v}\vec{v})-\nabla p +\rho\vec{g}+\vec{F}_{\rm{SR}}+\vec{F}_{\rm{L}},\\
        \begin{split}
                \partial_{t}(\epsilon_{h}+\rho v^2/2)=&-\nabla\cdot(\vec{v}(\epsilon_{h}+p+\rho v^2/2))+\\&\vec{v}\cdot(\vec{g}+\vec{F}_{\rm{L}}+\vec{F}_{\rm{SR}} )+Q_{\rm{rad}}+Q_{\rm{res}}
        \end{split},\\
        \partial_{t}\vec{B}=&\nabla\times(\vec{v}\times\vec{B}).
\end{align}

Here, $\vec{F}$ refers to forces and $Q$ refers to source terms. The subscript $\rm{SR}$ refers to semi-relativistic "Boris correction"-related terms \citep{boris1970,gombosi2002}, which are negligible for our setups, and $\rm{L}$ refers to the Lorentz force. The two $Q$ terms in the energy equation account for the radiative heating/cooling and resistive heating (since the hydrodynamic energy is conserved, instead of the total energy). For details, we refer the reader to \citet{rempel2014,rempel2017}. In this work, the grey approximation is used for solving the radiative transfer equations, where the frequency dependence of the opacity is replaced by an average value. This is an acceptable approximation for this work since we are mainly interested in the structure below and just above the surface \citep{voegler2004grey}. Finally, the FreeEOS equation of state \citep{freeeos} with solar abundances \citep{asplund2009} is used to close the set of equations.

The effective temperature ($T_{\mathrm{eff}}$) (related to the radiative output), the surface gravitational acceleration ($g$), related to the hydrostatic balance, and the metallicity ($Z$) together uniquely specify the spectral class of a star. The \textit{MURaM} code uses a constant gravitational acceleration $g$, the gas pressure at bottom boundary ($p_{\mathrm{bot}}$), and the inflow entropy at the bottom boundary ($s_{\mathrm{bot}}$) as free parameters. The $p_{\mathrm{bot}}$ and $s_{\mathrm{bot}}$ parameters determine the height of the $\tau=1$ surface and the $T_{\mathrm{eff}}$.

\subsection{Setup and parameters}\label{sec:methods:setup}

We considered four stellar cases: F3V, G2V, K0V, and M0V,  chosen to cover a broad range of $T_{\mathrm{eff}}$ for stellar types with convective envelopes. All boxes have the same number of grid-points (512 $\times$ 512 in the horizontal direction and 500 in the vertical direction). The scaling for the horizontal and the vertical geometric extent was done such that the number of granules in each box is roughly the same and the number of pressure scale heights below the photosphere is also similar ($\sim 7.5$). For the G-star, this corresponds to 4 Mm below the surface and a horizontal extent of 9 Mm $\times$ 9 Mm.

\begin{table*}[h]
    \caption{Parameters for the simulation setup}
    \centering
    \begin{tabular}{c | c c c c | c c c c c c}
    \hline\hline
        Type & $z_{\downarrow}$\tablefootmark{*}$(z_{\mathrm{tot}})$ & $x_0,y_0$ & $\Delta x,y$ & $\Delta z$ & $\log_{10} g$ & $(T_{\mathrm{eff}})_{HD}$\tablefootmark{*} &  $(T_{\mathrm{eff}})_{\mathrm{SSD}}$\tablefootmark{*} & $\avg{||B||}_{\tau=1}$\tablefootmark{*} & $\avg{|B_{z}|}_{\tau=1}$\tablefootmark{*} & $\avg{||B||}_{\avg{\tau}=1}$\tablefootmark{*} \\
         & (Mm) & (Mm) & (km) & (km) & (cm/s$^{2}$) & (K) & (K) & (G) & (G) & (G) \\
    \hline
        F3V & 11.11 (13.00) & 23 & 45 & 26 & 4.301 & 6817$\pm$7 & 6807$\pm$8 & 188$\pm$15 & 93$\pm$8 & 132$\pm$10 \\
        G2V & 4.09 (5.00) & 9 & 17.5 & 10 & 4.438 & 5834$\pm$9 & 5840$\pm$9 & 127$\pm$12 & 66$\pm$6 & 113$\pm$10\\
        K0V & 2.05 (2.31) & 4.62 & 8.2 & 4.62 & 4.609 & 4668$\pm$5 & 4671$\pm$5 & 103$\pm$6 & 50$\pm$3 & 103$\pm$6\\
        M0V & 0.90 (1.14) & 2.043 & 3.99 & 2.28 & 4.826 & 3825$\pm$1 & 3827$\pm$2 & 106$\pm$6 & 54$\pm$4 & 107$\pm$6\\
            \hline
    \end{tabular}
    \label{table:params}
    \tablefoot{
    \tablefoottext{*}{These quantities are determined after running the simulations. The change in $T_{\mathrm{eff}}$ will not influence the total radiative output on long timescales ($>10^5$ yr) corresponding to the Kelvin-Helmholtz timescale. See \citet{spruit1982} for details.}
    }
\end{table*}

The boundaries are periodic in the horizontal $x,y$ direction. The top boundary ($z_{\mathrm{top}}$) is open to outflows and closed to inflows, with vertical magnetic fields. The bottom boundary ($z_{\mathrm{bot}}$) is symmetric\footnote{This refers to the way the derivative is handled across the ghost cells. Symmetric boundary implies the same value in the ghost cell ($q_g$) next to the boundary domain cell ($q_d$), such that the derivative across the boundary is zero ($q_g=q_d$), and anti-symmetric implies a value with the opposite sign ($q_g=-q_d$)} for mass flux ($\rho\vec{v}$), entropy downflows, and magnetic fields. This magnetic field boundary condition also allows horizontal field to be advected across the bottom boundary. This "mimics" the presence of magnetic field deeper in the convection zone, as considered previously on the basis  of equipartition arguments \citep{rempel2014,hotta_ssd}. The magnetic field BC may not necessarily preserve the $\nabla \cdot \vec{B}=0$ constraint. However, the hyperbolic divergence cleaning approach \citep{divbclean} employed in \textit{MURaM} takes care of the $\nabla \cdot \vec{B}$ errors reasonably well: $(\nabla\cdot\vec{B})_{\mathrm{rms}}/(||B||/\Delta z) \sim O(10^{-3})$; we refer to Fig. \ref{fig:divB} for details.

For each star, we performed simulations with magnetic fields (SSD), and purely hydrodynamic (HD) simulations. HD simulations were initialized with density and internal energy (IE) profiles generated using the 1D stellar code MESA \citep{Paxton2019} for the F-, K-, and M-star and using the standard solar model from \citet{cd1996} for the G-star. These were then run for several hours in stellar time till convection became relaxed and there were no transients visible in velocity and pressure vertical slices. 
Then the simulation box was seeded with net zero-flux magnetic field with a negligibly small field strength ($10^{-5}$ G) and run till photospheric magnetic field strength reached saturation. The results presented in the subsequent sections are averaged over a few hours of stellar time (after saturation), and over a number of snapshots, for all the eight cases (see Table \ref{tab:color_code} for further details for each setup).

Table \ref{table:params} describes the detailed setup for all the simulations: for all four stellar types, it gives the height of the $\tau=1$ surface above the bottom boundary (and the total vertical extent), the horizontal extent, the horizontal resolution, the vertical resolution, the log constant surface gravity, the effective temperature of the SSD and HD cases, the average magnetic field magnitude at the $\tau=1$ iso-surface\footnote{This is the surface corresponding to where $\tau=1$ in each vertical column of the simulation cube, and this is usually somewhat corrugated because downflows are cooler than upflows and opacity is extremely sensitive to temperature in the relevant ranges.} $\avg{||B||}_{\tau=1}$, the average unsigned vertical field at the $\tau=1$ iso-surface $\avg{|B_z|}_{\tau=1}$, and the average magnetic field magnitude at the $\avg{\tau}=1$ horizontal slice $\avg{||B||}_{\avg{\tau}=1}$ of the SSD cases. The effective temperature is calculated by averaging the angle-averaged bolometric luminosity over time.
\section{Results}\label{sec:results}

\begin{table}[h]
    \caption{Color-coding for the plots, along with number of snapshots considered and total time in hours for each stellar box}
    \label{tab:color_code}
    \centering
    \begin{tabular}{c | c c c | c c c}
        \hline\hline
        Star & & SSD &  & & HD &  \\
         & Color & $N$ & $t$ (h) & Color & $N$ & $t$ (h) \\
        \hline
        F3V & \textcolor{blue}{Blue} & 264 & 12.4 & \textcolor{dodgerblue}{Light Blue} & 124 & 14.8 \\
        G2V & Black & 178 & 7.7 & \textcolor{gray}{Gray} & 117 & 15.0 \\
        K0V & \textcolor{darkgreen}{Green} & 125 & 10.6 & \textcolor{green}{Lime} & 193 & 16.8 \\
        M0V & \textcolor{red}{Red} & 263 & 5.4 & \textcolor{indianred}{Light Red} & 197 & 10.1\\
        \hline
    \end{tabular}
    \tablefoot{All HD plots are dashed lines.}
\end{table}

\begin{figure*}[h]
\centering
\includegraphics[width=17cm]{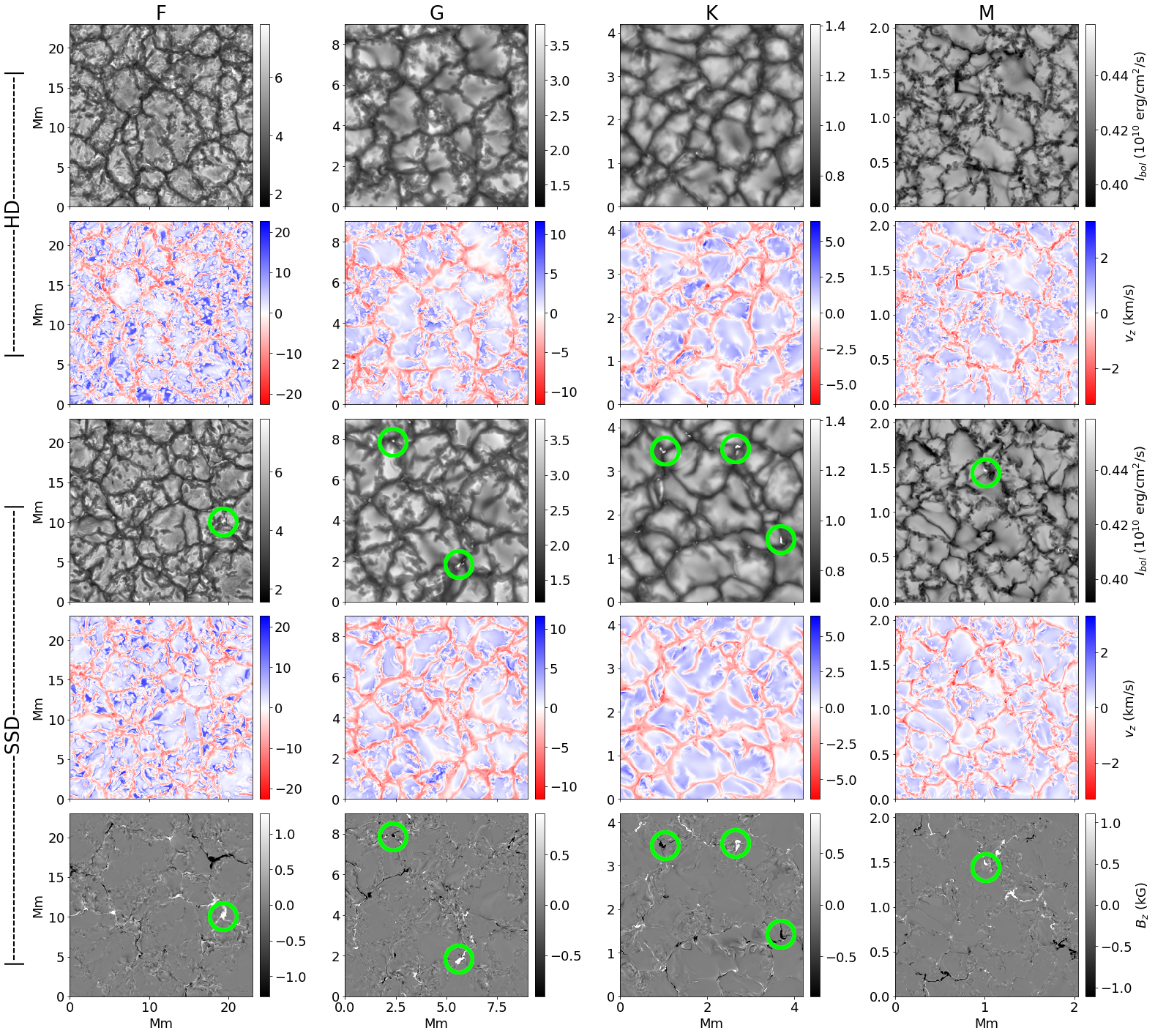}
\caption{Emergent intensity and surface vertical velocities in different stellar types for models with and without magnetic field. \textit{From top to bottom:} Snapshot of the bolometric intensity and $v_z$ at $\tau=1$ for the HD case (\textit{rows 1 and 2}), bolometric intensity and $v_z$ at $\tau=1$ for the SSD case (\textit{row 3 and 4}), and the corresponding vertical magnetic field at $\tau=1$ (\textit{row 5, from left to right}) for spectral types F, G, K, and M, respectively. The green circles indicate the bright points and corresponding magnetic field concentrations.}
\label{fig:snap}
\end{figure*}

All the magnetic simulations exhibit dynamo action and develop magnetic fields with energy within an order of magnitude of the kinetic energy (KE) through most of the simulation box. The change in the partition of energy influences the internal structure as well as convective velocities. Snapshots of the bolometric intensity and vertical velocity at the $\tau=1$ surface for the SSD and the HD setups (along with the vertical magnetic field for the SSD cases) are shown in Fig. \ref{fig:snap}. The SSD cases show distinct intergranular bright points which correspond with strong magnetic field concentrations. In the following subsections, we examine the horizontally averaged structure of the magnetic field and its effects on the stratification as well as convection for these stellar types. The analysis of the magnetic fields in the lower photosphere and their effect on the bolometric intensity and vertical velocity will be covered in the next paper in this series.

All 1D plots are averages over a number of snapshots spanning a few hours of stellar time (see Table \ref{tab:color_code} for exact numbers). The error bars are standard error (standard deviation normalized by the square root of the number of snapshots $\sigma/\sqrt{N}$) of the average 1D structure, the assumption being that over this time span, the snapshots are statistically independent.

All quantities are plotted as a function of number of pressure scale heights relative to the height where $\avg{\tau}=1$, $n_H = \log(p_{\mathrm{gas}}/p_{\mathrm{gas}(\avg{\tau}=1)}).  $ We note that with this definition, positive values correspond to the interior. Since the non-magnetic bottom boundary conditions are identical ($p_{\mathrm{bot}}$ and $s_{\mathrm{bot}}$ are the same) for HD and SSD runs, the deviations are calculated geometrically and plotted against the corresponding HD pressure scale axis.

This extent in terms of pressure scale heights ranges from 7.5 (bottom) to -5 (near top) for all the simulations. Since the focus of this paper is on the near surface convection zone, we excluded  the portion of the box corresponding to $n_H < -1 $ from our analysis. We also excluded the region corresponding to $n_H > 6$ due to possible numerical bottom boundary effects.

\subsection{Magnetic field structure}\label{sec:results:mag}

\begin{figure}[h]
    \resizebox{\hsize}{!}
    {\includegraphics{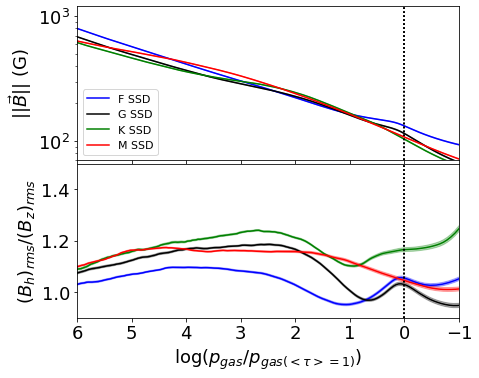}}
    \caption{\textit{Magnetic field structure for the four stellar cases. \textit{Top:} Horizontally averaged magnetic field magnitude. \textit{Bottom:} The ratio of the horizontal rms field strength to the vertical r.m.s field strength. The horizontal axis is the number of pressure scale heights $\log_{10}(p_{\mathrm{gas}}/p_{\mathrm{gas}(\tau=1)})$, calculated for the HD cases, below the surface (\textit{dotted vertical black line}). The shaded regions correspond to 1-$\sigma$ standard error ($e=\sigma/\sqrt{N}$, $N$ is the number of snapshots) computed over time averaging of snapshots.}}
    \label{fig:mag_str}
\end{figure}

As mentioned in the introduction to this section, the dynamo action results in a significant amount of magnetic field, with the overall magnitude  roughly similar for all the cases, and with a somewhat decreasing trend with $T_{\mathrm{eff}}$ near the surface (Fig. \ref{fig:mag_str}, top panel and last column of Table \ref{table:params}). The relation of the magnetic energy (ME) to kinetic energy is discussed in §\ref{sec:results:ener}.

The bottom panel of Fig. \ref{fig:mag_str} shows the ratio $B_h/B_z$, which gives an indication of the 3D structure of the magnetic fields. For a fully isotropic distribution of magnetic field, one would expect $B_x^2 \approx B_y^2 \approx B_z^2$. This implies that the ratio of the horizontal r.m.s. component of the magnetic field fluctuations $B_{\mathrm{h,rms}}$ to the vertical component $B_{\mathrm{z,rms}}$ should be $\sim \sqrt{2}$. For G-, K-, and M-stars, this ratio is slightly less than $\sqrt{2}$ in the middle of the box ($2<n_H<5$), indicating near-isotropy, whereas for the F-star, it is significantly lower ($\sim$ 1). 

Near the $\tau=1$ surface, $B_{\mathrm{h,rms}}/B_{\mathrm{z,rms}}$ for all the stars is lower because of intensification of vertical magnetic fields in the intergranular lanes (e.g., \citet{spruit1979}).

\subsection{Changes in stratification}\label{sec:results:strat}

\begin{figure}[h]
    \resizebox{\hsize}{!}
    {\includegraphics{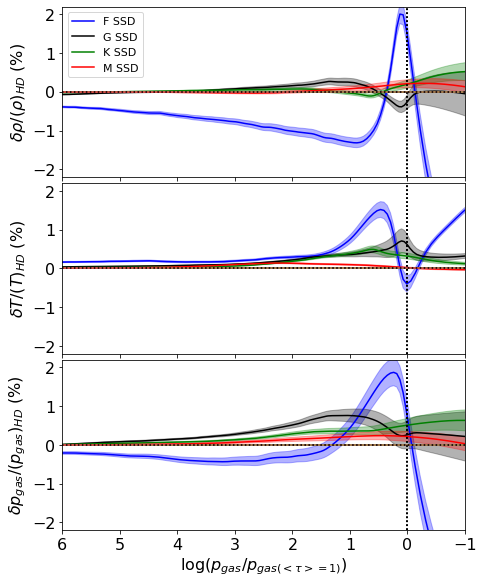}}
    \caption{Deviations in $\rho$, $T,$ and $p_{\mathrm{gas}}$ for F (\textit{blue}), G (\textit{black}), K (\textit{green}), and M-star (\textit{red}) cases. The vertical axis gives the geometric deviations as a percentage relative to the hydrodynamic case. The horizontal axis is the number of pressure scale heights $\log_{10}(p_{\mathrm{gas}}/p_{\mathrm{gas}(\tau=1)})$, calculated for the HD cases, below the surface (\textit{dotted vertical black line}). The shaded regions correspond to 1-$\sigma$ standard error ($e=\sigma/\sqrt{N}$, $N$ is the number of snapshots) of the mean solid curve.}
    \label{fig:td_dev}
\end{figure}

Figure \ref{fig:td_dev} shows the plots of deviation from the mean HD stratification. The deviations from the HD simulations are presented as the relative percent change in the horizontally averaged 1D structure. For any quantity of interest $q$ (e.g., density, temperature, etc.), these deviations are calculated as $(q_{\mathrm{SSD}}-q_{\mathrm{HD}})/q_{\mathrm{HD}}$. This means that a positive value for the deviation corresponds to a higher value for the SSD case relative to the HD case.

All simulations show slight ($\leq$ 2\%) changes in thermodynamic stratification relative to the corresponding HD simulations, with the magnitude of deviations below the surface roughly increasing from coolest to hottest stellar type. Below 3 pressure scale heights, the G, K, and M simulations show negligible $(\leq 0.1 \%)$ deviations in thermodynamic quantities. 

Closer to the surface, the F-star simulation shows up to 1.5\% reduction in density and up to 1\% reduction in gas pressure (Fig. \ref{fig:td_dev}, top and bottom panels, blue line). This trend is opposite to that seen for other cases, all of which show a slight ($<1\%$) enhancement in density and pressure. These results are analyzed in §\ref{sec:discuss:pturb}. For context, a 1.5\% deviation in temperature for an F-star would correspond to a temperature change of $\sim$100 K.

\subsection{Distribution of energies}\label{sec:results:ener}

\begin{figure}[h]
    \resizebox{\hsize}{!}
    {\includegraphics{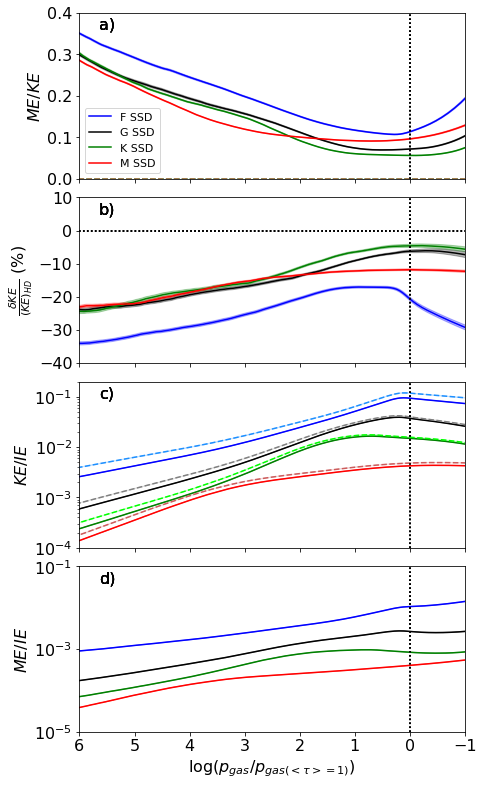}}
    \caption{Distribution of energies for the four stellar cases. a) Ratio of ME to KE, b) percent change in KE, c) ratio of KE to IE, and d) ratio of ME to IE (\textit{bottom}) for F-, G-, K-, and M-stars.}
    \label{fig:ener_char}
\end{figure}

In Fig. \ref{fig:ener_char}, panel a shows the ratio of magnetic to kinetic energy. For all cases, this ratio is within an order of magnitude throughout the box. This ratio for the F-star is significantly higher compared to the other stars. In addition, the ratio for M-star has a minima deeper down in the box than for other stars. As before, the general trend shows a decrease in this ratio with decreasing $T_{\mathrm{eff}}$.

In Fig. \ref{fig:ener_char}b, we shows the relative change in KE between the SSD and the HD setups. All stars show a marked decrease in KE in the SSD case. As with the top panel, F-star shows a significantly stronger reduction in KE compared to the other cases. The M-star as well shows a stronger reduction in KE near the surface compared to the G and K-stars.

Fig. \ref{fig:ener_char}c shows the ratio of kinetic to internal energy. For the F-star, this ratio is within an order of magnitude near the surface, whereas for the other stars, it is less than $1\%$.
Fig. \ref{fig:ener_char}d shows the ratio of the magnetic to the internal energy. There is a clear trend in this ratio with stellar type, with it being lowest for M-star ($\sim 5\times10^{-5}$ near bottom, to $\sim 10^{-4}$ near surface) and highest for F-star ($\sim 10^{-3}$ near bottom, to $\sim 10^{-2}$ near surface).

\subsection{Changes in velocities}\label{sec:results:vel}

\begin{figure}[h]
    \resizebox{\hsize}{!}
    {\includegraphics{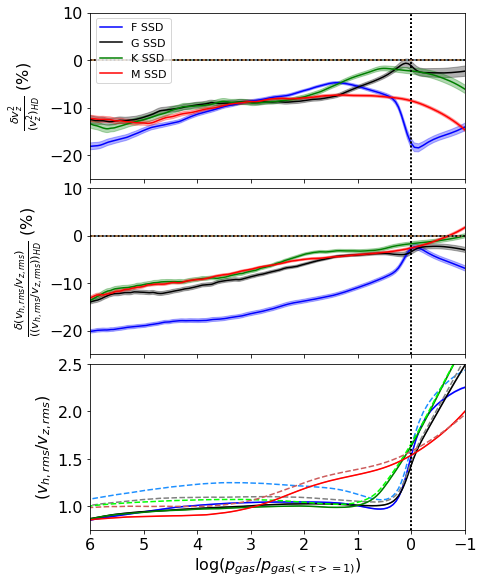}}
    \caption{Relative decrease in convective velocities $v_z^2$ (\textit{top}), relative decrease in ratio of horizontal to vertical rms velocities $v_{\mathrm{h,rms}}/v_{\mathrm{z,rms}}$ (\textit{middle}) and the actual $v_{\mathrm{h,rms}}/v_{\mathrm{z,rms}}$ ratio (\textit{bottom}).}
    \label{fig:vels}
\end{figure}

All SSD cases exhibit a decrease in vertical velocities $v_z^2$ as well as the ratio of horizontal to vertical rms velocities $v_{\mathrm{h,rms}}/v_{\mathrm{z,rms}}$ (Fig. \ref{fig:vels}), relative to the corresponding HD cases. The reduction in $v_z^2$ follows similar trend for all the four stars, with a decrease of 4-8\% near the surface, going up to 20\% near the bottom boundary. For G-, K- and M-star, the decrease in $v_{\mathrm{h,rms}}/v_{\mathrm{z,rms}}$ is similar (5\% near the surface, going up to 12\% near the bottom) but is more pronounced for the F-star case (10\% near the surface, going up to 20\% near the bottom boundary). This implies a change in the horizontal extent of sub-surface granulation, which follows \citet{nordlund2009}, who showed using simple mass conservation that the horizontal extent of granules is proportional to $H(v_h/v_z)$, where $H$ is the local density scale height.

The ratio $v_{\mathrm{h,rms}}/v_{\mathrm{z,rms}}$ gives an idea of the 3D velocity structure. Close to the $\tau=1$ surface, the ratio increases suddenly. This increase corresponds to where convective flows turn over, as the atmosphere becomes convectively stable and the flows above that point are mainly due to convective overshoot. The exact pressure scale depth where this turning over takes place depends of the effective temperature: for the F- and G-star, this turnover takes place within half a pressure scale height of the $\tau=1$ surface whereas, for the M-star, it takes place well below the $\tau=1$ surface (around $n_H=3$). This is also probably why the minima of the ME/KE ratio in Fig. \ref{fig:ener_char} for M-star is significantly below the surface compared to the G and K-star, as density (and, consequently, KE) is lower above the height where most of the overturning takes place.
The changes in $v_z^2$ as well as the $v_{\mathrm{h,rms}}/v_{\mathrm{z,rms}}$ with depth seem to follow a regular trend, and is consistent with the results obtained previously in whole-convection zone simulations with a small-scale dynamo setup \citep{hotta_ssd}.
\section{Discussion}\label{sec:discuss}
\subsection{Turbulent pressure}\label{sec:discuss:pturb}

\begin{figure}[h]
    \resizebox{\hsize}{!}
    {\includegraphics{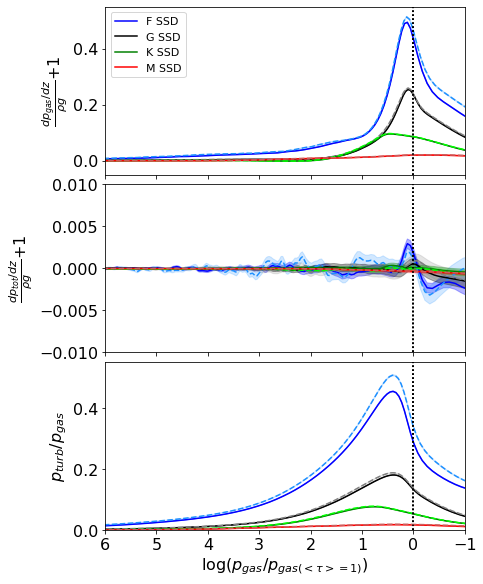}}
    \caption{Hydrostatic force balance with gas pressure $p_{\mathrm{gas}}$ (\textit{top}) and total pressure $p_{\mathrm{tot}}$ (\textit{middle}) gradient term and $\rho g$ term, normalized by $\avg{\rho g}$ as $\avg{d p/dz + \rho g}/\avg{\rho g}$, and the ratio of turbulent pressure to gas pressure (\textit{bottom}). All quantities are plotted for SSD (\textit{solid}) and HD (\textit{dashed}) cases. We note that the extent of $y$-axis is larger by more than an order of magnitude for force balance with $p_{\mathrm{gas}}$ compared to force balance with $p_{\mathrm{tot}}$}
    \label{fig:fbal}
\end{figure}

We first discuss the changes in the pressure and density in the SSD models relative to a purely HD model. The pressure and density changes (Fig. \ref{fig:td_dev}) are very small ($\sim 0.1\%$) for the G-, K-, and M-star in the convection zone (and not significantly greater near the surface for the K- and M-star). The changes are more prominent for the F-star ($\sim 1\%$). This discrepancy can be understood by considering the contribution of turbulent pressure in the overall hydrostatic balance. 

Turbulent pressure becomes important for hydrostatic balance when KE is within an order of magnitude or so of the IE. This is indeed the case for our F-star models. This implies that turbulent pressure (which is largely due to plasma motions) can be a significant fraction of the gas pressure $p_{\mathrm{gas}}$. Quantitatively, this can be seen from a crude MLT calculation of the Mach number (see appendix \ref{app:mach} for derivation):
\begin{align}\label{eqn:mach}
        M & \approx 0.138 \left(\frac{T}{\text{10$^{3}$ K}}\right)^{5/6} \left(\frac{\rho}{\text{10$^{-7}$ g/cm$^{3}$}}\right)^{-1/3} .
\end{align}
Using the above equation, we calculate the F-star photospheric Mach number to be about 0.75 (see Table \ref{tab:mach}), slightly lower than simulation value in \citet{beeck1} who found it to be 0.9. For the other stars, the velocity is decidedly subsonic.

\begin{table}[h]
    \caption{Mach numbers computed using MLT}
    \label{tab:mach}
    \centering
    \begin{tabular}{c c c c}
        \hline\hline
        Simulation & $T_{\mathrm{surf}}$ (10$^3$ K) & $\rho_{\mathrm{surf}}$ (10$^{-7}$ g/cm$^3$) & $M$ \\
        \hline
        F3V & 6.19 & 0.59 & 0.75\\
        G2V & 6.17 & 2.39 & 0.47\\
        K0V & 4.94 & 6.67 & 0.28\\
        M0V & 3.96 & 21.3 & 0.16\\
        \hline
    \end{tabular}
    \tablefoot{$\rho_{\mathrm{surf}}$ and $T_{\mathrm{surf}}$ are obtained from HD simulation data}
\end{table}

Now, the hydrostatic balance is expressed in terms of the balance between force due to pressure gradient, $p' = dp/dz,$ and gravity, $\rho g$, acting in the vertical (viz. radial) direction. These two terms should be approximately equal in magnitude. If just $p_{\mathrm{gas}}$ is considered, this balance does not hold very well, with deviations increasing strongly with $T_{\mathrm{eff}}$ (Fig. \ref{fig:fbal}, top panel).

The turbulent pressure consists of terms from the total stress tensor. From Reynolds stresses, the $\rho v_i v_j$ term and from Maxwell stresses, the $((B^2/2)\delta_{ij} - B_i B_j)/4\pi$ term is obtained. As mentioned in the introduction, the presence of SSD magnetic fields implies a reduction in KE, where energy is redistributed between the plasma motions and magnetic fields via Lorentz force feedback. In Fig. \ref{fig:ener_char}, the ratio of ME to KE (top panel) corresponds remarkably well with the reduction in KE relative to the HD case (middle panel), implying most of the energy in the magnetic fields is drawn from the KE reservoir. The ME is within an order of magnitude of the KE for subsurface plasma. This is consistent with the results on equipartition of energy in several SSD simulations  \citep{hotta_ssd,haugen_ssd,schekochihin_ssd}.

This order of magnitude equipartition results in a reduction of plasma velocities and, consequently, the magnitude of the Reynolds stress term. The contribution from the Maxwell stress, on the other hand, can be either negative or positive, depending on whether the effects of magnetic tension dominate over that of magnetic pressure in the vertical direction. More quantitatively (see Appendix \ref{app:turb} for a derivation), the total turbulent pressure can be expressed as:

\begin{align}\label{eqn:pturb}
    p_{\mathrm{turb}}&=\rho v_z^2 + \frac{B_h^2-B_z^2}{8\pi}.
\end{align}
With this included in the calculation of the pressure gradient term, the hydrostatic balance is satisfied, as can be seen in middle panel of Fig. \ref{fig:fbal} (noting the difference in the extent of $y$-axis for the top and middle panels). In the bottom panel of the same figure, the reduction in $p_{\mathrm{turb}}/p_{\mathrm{gas}}$ is most prominent for the F-star. Since the changes in $v_z^2$ are similar (Fig. \ref{fig:vels}, top panel) for all stars, there must be a reduction in density for the F-star to compensate for the significant change in $p_{turb}$.
Based on the expression for $p_{\mathrm{turb}}$, we introduce an effective turbulent velocity $v=\sqrt{p_{\mathrm{turb}}/\rho}$.

With this, it becomes possible to relate changes in density stratification to changes in total pressure gradient. A straightforward consideration of the density scale height $H_{\rho}$ (see appendix \ref{app:turb}) yields:
\begin{align} \label{eqn:rho_scale}
    H_\rho &=RT/(\mu g) + v^2/g.
\end{align}

This tells us that where $B_h^2 < B_z^2$ and $(\rho v_z^2)_{\mathrm{SSD}} < (\rho v_z^2)_{\mathrm{HD}}$, the scale height of the SSD model is smaller than for the HD model, $(H_{\rho})_{\mathrm{SSD}} < (H_{\rho})_{\mathrm{HD}}$. This is valid if assuming that the change in $T$ and $\mu$ is relatively small. From the proportionality between density scale height and $v_z^2$ (Eq. \ref{eqn:rho_scale}), the decrease in vertical velocity is associated with a decrease in the local density scale height. This can be inferred from Fig. \ref{fig:fbal} (bottom panel), where the ratio $p_{\mathrm{turb}}/p_{\mathrm{gas}}$ ($\propto v^2/(RT/\mu)$) is noticeably lower for F-star near the surface.

These conditions are satisfied relatively well for the F-star case. A decrease in $H_{\rho}$ implies a steepening of the density stratification for the SSD case relative to the HD case. This is exactly the case in the top panel of Fig. \ref{fig:td_dev}: the reduction in density goes from a $\sim$0\% to $\sim1.5$\% near the surface. Below the surface, it is reasonably good to assume a perfect gas equation of state, with $p_{\mathrm{gas}} \propto \rho T$. Assuming changes in $T$ are small compared to $\rho$ and $p_{\mathrm{gas}}$, this, in turn, implies that the changes in $p_{\mathrm{gas}}$ follows a similar trend as changes in $\rho$. Near the surface, this picture breaks down since the mode of energy transfer changes from convective to radiative and this simplified analysis is no longer valid.

The effect of turbulent pressure on stratification and convection has previously been considered for stellar model envelopes from an MLT perspective \citep{henyey1965} as well as for 3D HD simulations \citep{ludwig2012,jorgensen2019}; however, to the best of our knowledge, the effect of magnetic fields from an SSD have not been considered before.

\subsection{Changes in velocity structure}\label{sec:discuss:magvel}

As mentioned in §\ref{sec:results:vel}, the ratio $v_{\mathrm{h,rms}}/v_{\mathrm{z,rms}}$ (Fig. \ref{fig:vels}, bottom panel), gives an idea of where convection overturns relative to pressure scale height for a given effective temperature. The trend in overturning for different spectral types fits the usual picture of "hidden" granulation below the $\tau=1$ surface for cooler, denser stars such as M and K versus the "naked" granulation above the $\tau=1$ surface of hotter, more rarefied stars such as G and F, as discussed in \citet{nordlund1990granulation}.

This ratio is close to 1 for the HD G-, K-, and M-stars (dashed lines). For the HD F-star, however, it is significantly higher (dashed light blue line), indicating a higher degree of isotropy in the velocity structure\footnote{Fully isotropic flow requires $v_x^2=v_y^2=v_z^2$. Hence, this would imply $v_{\mathrm{h,rms}}/v_{\mathrm{z,rms}}=\sqrt{2}$}. This implies relatively higher horizontal velocities for the HD F-star. The near-isotropic velocity profile for the HD F-star can be attributed to the KE being a non-negligible fraction of the IE. We speculate that these stronger horizontal velocities contribute to the stronger magnetic fields in the vertical direction for the F-star (as inferred from the lower $B_{\mathrm{h,rms}}/B_{\mathrm{z,rms}}$ for the F-star in Fig. \ref{fig:mag_str}, bottom panel).

In the case of SSD (solid lines), all stars follow a similar trend for $v_{\mathrm{h,rms}}/v_{\mathrm{z,rms}}$, with the ratio being $\leq 1$ up to the point where convection overturns. The change in velocity field structure in the presence of SSD fields is a hard problem that depends on the SSD saturation mechanism. We note that the $B_{\mathrm{h,rms}}/B_{\mathrm{z,rms}}$ ratio is lower for F-star compared to other cases. It is plausible that the relatively stronger fields in the vertical direction restrict horizontal flows more in the case of F-stars and lead to a greater change in the value of $v_{\mathrm{h,rms}}/v_{\mathrm{z,rms}}$. However, a fuller understanding of this behavior requires a more detailed analysis that is beyond the scope of this paper.
\section{Conclusions}\label{sec:conclude}

In this work, we  investigate the magnetic field self-consistently generated by an SSD acting in the near-surface layers of main-sequence stars of spectral types F3V, G2V, K0V, and M0V. The SSD mechanism operates in all cases to amplify magnetic fields from a seed field of negligible strength and zero net flux. The magnetic fields from the SSD have an energy density that is a non-negligible fraction of the kinetic energy density. These fields act back on the plasma to reduce the convective velocities, which in turn reduces the turbulent pressure. This becomes substantial for the F-star as it is hot enough to have kinetic and internal energy within an order of magnitude near the surface, which gives magnetic fields stronger than those in G-, K-, and M-stars, especially in the vertical direction. The equation for hydrostatic balance for total pressure and the reduction of convective velocities implies a reduction in the density scale height itself. This is significant enough for the F-star to result in reduced density and gas pressure throughout the box. This effect tends to get smaller towards later spectral types.

This paper only covers the near-surface convection zone. Other aspects of particular interest to observational studies would include the magnetic field structure in the lower photosphere and changes in the intensity characteristics. In addition, the change in scale height and the changes in $v_h/v_z$ imply changes in granulation scale. All these points are to be investigated in a follow-up paper.

\begin{acknowledgements}
    TB ran the simulations, performed the analysis and wrote most of the text. DP helped with running the simulations. All authors contributed to the final version of the text. We thank the anonymous referee for their helpful and encouraging comments. TB is grateful for access to the supercomputer Cobra at Max Planck Computing and Data Facility (MPCDF), on which all the simulations were carried out. We also acknowledge Alan Irwin for providing the open source FreeEOS equation of state. This project has received funding from the European Research Council (ERC) under the European Union’s Horizon 2020 research and innovation programme (grant agreement No. 695075 and 715947).
\end{acknowledgements}

\bibliographystyle{bibtex/aa} 
\bibliography{bibtex/biblio.bib} 

\begin{appendix}
\setlength{\parindent}{0em}
\section{Mach number derivation}\label{app:mach}
We derive the expression for Mach number used in Eq. (\ref{eqn:mach}) from §\ref{sec:discuss:pturb}. Considering the hydrostatic pressure balance equation, the expression for pressure-scale height ($H_{p}$), the expression for sound speed ($c_{s}$) and the ideal gas equation of state:
\begin{align}
        p'&=-g\rho \label{gas:hdbal},\\
        H_{p}&=-p/p' \label{gas:hp},\\
        c_{s}^{2}&=\gamma p/\rho \label{gas:cs},\\
        p&= \rho (R/\mu) T. \label{gas:eos}
\end{align}
Here, $H_p$ is the pressure scale height, $p'=dp/dz$, $\mu$ is the mean molar mass, and $1/\mu=(1+E)(X+Y/4+Z/2)$, where $E$ is ionization fraction and $X,Y,Z$ are the H, He and metal abundances.
In the B\"ohm-Vitense MLT \citep{bv_mixinglength}, convective flux and velocities are expressed as:
\begin{align}
        F_{conv}&={\alpha c_{p}} \rho v T (\nabla-\nabla_{a})/2 \label{mlt:F},\\
        v^{2}&={\alpha^{2}\delta} g H_{p} (\nabla-\nabla_{a})/8. \label{mlt:v2}
\end{align}
Here, $\alpha$ is the mixing length parameter (usually taken to be between 1.5 to 2), $\delta=1-(\partial \ln \mu/\partial \ln T)_{p}$ (abundance gradient with temperature for an ideal gas), $c_{p}=(\partial U/\partial T)_{p} + p\delta/(\rho T)$ (heat capacity at constant pressure), and $\nabla=d\ln{T}/d\ln{p}$, with the subscript $a$ referring to `adiabatic.' We refer to chapter 6 of \citet{stixsunbook} for more details.\\ 

Near the surface, as most energy is carried by radiation, $F_{conv}\approx \sigma T^4$. We take the ratio of Eq. \ref{mlt:F} and Eq. \ref{mlt:v2} to eliminate $\nabla-\nabla_a$ and use $\sigma T^4$ instead of $F_{conv}$. Taking Mach number $M=v/c_{s}$ and $c_{s}=\sqrt{\gamma R T/\mu}$ and eliminating $H_p$ and $p$ using Eq. \ref{gas:hdbal}, \ref{gas:hp}, and \ref{gas:eos},
\begin{align} \label{mach}
        M&=\left(\frac{\alpha \delta \sigma \mu^{1/2}}{4c_{p}\gamma^{3/2}R^{1/2}}\right)^{1/3}T^{5/6}\rho^{-1/3}.
\end{align}
Using relevant values for the terms in the parentheses (all in cgs), we get Eq. (\ref{eqn:mach}) in terms of $T$ and $\rho$ as above:
\begin{align*}
        \alpha &=1.8 & \text{(average literature value)}\\
        \delta &=1 & \text{(assume $\mu$ is constant)}\\
        \sigma &= \text{5.67e-5 erg/cm$^2$/s/K$^{4}$} & \text{(Stefan-Boltzmann constant)}\\
        \gamma &= 5/3 &\text{(monoatomic adiabatic index)}\\
        \mu &= 1.2 & \text{(solar surface abundance)}\\
        R &= \text{8.314e7 erg/K/mol}& \text{(universal gas constant)}\\
        c_{p} &= \gamma R/(\mu(\gamma-1)) & \text{(isobaric specific heat capacity)}
\end{align*}

The value of $\mu=1.2$ assumes $E \sim 0$ near $\tau=1$. For an M-star, this may not necessarily be the case. However, even if $E\to 1$, $\mu$ changes by a factor of 2 and, accordingly, $M$ changes by a factor of $2^{1/2} \approx 1.4,$ and the qualitative result still holds.

\section{Turbulent pressure and pressure scale height derivation}\label{app:turb}
In §\ref{sec:discuss:pturb}, we use the following expression for $p_{turb}$ in Eq. (\ref{eqn:pturb}). that can be derived by considering the total (Reynolds and Maxwell) stress tensor, $\sigma_{ij}$, for ideal MHD:
\begin{align}
    \sigma_{ij}&=\rho(v_i v_j)+\left(p+\frac{B^2}{8\pi}\right)\delta_{ij}-\frac{B_i B_j}{4\pi}.
\end{align}
Here $i,j$ represent the $x,y$ and $z$ directions.

The pressure balance along the vertical ($z$) direction involves the term $\avg{\nabla_z \cdot \sigma_{iz}}_z$ which is equal to $\partial p_{\mathrm{tot}}/\partial z$. We can now compare these two terms to get an expression for total pressure ($p_{tot}$). Assuming all the off-diagonal terms ($i\ne j$) are negligible (which is akin to saying there is no cross-correlation between the vertical and the horizontal velocity and magnetic field components), we obtain the following:
\begin{align}
    \avg{p_{\mathrm{tot}}}_z&=\biggl<{\rho v_z^2+\left(p+\frac{B_x^2 +B_y^2 + B_z^2}{8\pi}\right)-\frac{\left(B_z^2+\cancelto{0}{B_zB_x}+\cancelto{0}{B_zB_y}\right)}{4\pi}}\biggr>_z,\\
    \avg{p_{\mathrm{tot}}}_z&=\avg{\rho v_z^2}+\avg{p}_z+\frac{\avg{B_h^2-B_z^2}_z}{8\pi}.
\end{align}
The next step is to consider how using $p_{tot}$ instead of $p$ affects the density scale height, $H_{\rho}$. In Eq. (\ref{gas:hdbal}), we substitute expression for $p_{tot}$, instead of $p$, and use Eq. (\ref{gas:eos}) to eliminate $p$. With this, we obtain:
\begin{align}
    \frac{d}{dz} \left( \rho (RT/\mu + v^2) \right) &=-\rho g.
\end{align}
Here, we take $v^2$ to be $v_z^2+(B_h^2-B_z^2)/(8\pi \rho).$ Then, the above equation can be rearranged to obtain:
\begin{align}
    \frac{d\rho}{\rho} &= - \frac{g dz}{RT/\mu +v^2} - \frac{d(RT/\mu +v^2)}{RT/\mu +v^2}.
\end{align}
From this, we can obtain the expression for density scale height, $H_\rho$, as:
\begin{align}
    H_\rho &=RT/(\mu g) + v^2/g.
\end{align}
The second term above includes the contribution from the turbulent pressure. Since this term is almost always smaller for the SSD case relative to the HD case (mainly because the convective velocities are lower for SSD), the corresponding pressure scale height for the SSD cases is also almost always smaller.

\section{Diagnostics}
For any MHD simulation, the computation and evolution of the magnetic field must be divergence free. To ensure this, \textit{MURaM} uses a hyperbolic divB cleaning algorithm \citep{divbclean}. Here, we show the horizontally averaged divergence of magnetic field across the box for all four SSD cases. Since the units for $\nabla \cdot B$ are field/length, a proper comparison requires a normalization. We do so with $||B||/\Delta z$. As noted in §\ref{sec:methods:setup}, at best, the error is $O(10^{-3})$.
\begin{figure}[h]
\resizebox{\hsize}{!}{\includegraphics{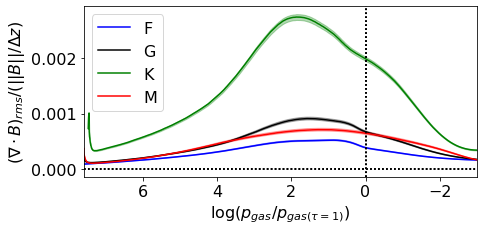}}
\caption{\textbf{div$\cdot$B} error for the four SSD cases: F, G, K, and M. The quantity plotted is horizontally averaged $(\nabla\cdot\vec{B})_{\mathrm{rms}}/(||B||/\Delta z)$.}
\label{fig:divB}
\end{figure}

\section{ERRATA to derivation of turbulent pressure in Appendix \ref{app:turb}}

The derivation of turbulent pressure contains a typo. The result is however correct, and the typo does not otherwise affect the paper. The first line in the right hand column on page 9 should have read 'The pressure balance along the vertical ($z$) direction involves the term $\langle \nabla_i \cdot \sigma_{iz}\rangle_z$ which is equal to $\partial P_\mathrm{tot}/\partial z$." Here, summation over $i$ is implied. The original sentence inadvertently had $\langle \nabla_z \cdot \sigma_{iz}\rangle_z$. The result then follows trivially from the fundamental theorem of calculus and the periodicity of ${\bf B}$ in the horizontal direction, which imply $\langle \frac{\partial B_x B_z}{\partial x} \rangle_z=\langle \frac{\partial B_y B_z}{\partial y} \rangle_z=0$.

\end{appendix}

\end{document}